\documentstyle[12pt]{article}
\def\be{\begin{equation}}
\def\ee{\end{equation}}
\def\ba{\begin{eqnarray}}
\def\ea{\end{eqnarray}}

\def\fun#1#2{\lower3.6pt\vbox{\baselineskip0pt\lineskip.9pt

\ialign{$\mathsurround=0pt#1\hfill##\hfil$\crcr#2\crcr\sim\crcr}}}
\begin{document}

\begin{titlepage}
\null\vspace{-62pt}
\begin{flushright} hep-ph/9604445 \\
UMD-PP-96-95\\
\today
\end{flushright}
\vspace{0.2in}

\centerline{{\Large \bf  A Supersymmetric Solution}}
 
\centerline{{\Large \bf  to {\it CP} Problems} }

\vspace{0.5in}
\centerline{\large \bf 
Rabindra N. Mohapatra\ \ and \ \ Andrija Ra\v{s}in }
\vspace{0.2in}
\centerline{\it Department of Physics}
\centerline{\it University of Maryland}
\centerline{\it College Park, MD 20742}
\vspace{.7in}
\baselineskip=19pt

\centerline{\Large \bf Abstract}
\begin{quotation}

We analyze the minimal supersymmetric left-right model with 
non-re\-nor\-malizable interactions induced by higher scale physics and 
study its {\it CP} violating properties. We show that it: 
(i) solves the strong {\it CP} problem;
(ii) predicts the neutron electric dipole moment well within
experimental limits (thus solving the usual SUSY {\it CP} problem).
In addition, it automatically conserves {\it R}-parity. 
The key points are that the parity symmetry forces the Yukawa 
couplings to be hermitean, while supersymmetry ensures that the scalar
potential has a minimum with real higgs doublet vacuum expectation values.
Gluino and B-L gaugino masses are automatically real. 
The observed {\it CP} violation in the kaon system comes, as in the Standard
Model, from the Kobayashi-Maskawa-type phases.
These solutions are valid for any value of the right-handed breaking 
scale $M_R$, as long as the effective theory below $M_R$ has only two Higgs 
doublets that couple fully to fermions. ({\it i.e.} the theory below $M_R$ is
MSSM-like.) 
The potentially dangerous $SU(2)_L$ gaugino one-loop contributions to 
$\bar{\Theta}$ below $M_R$ can be avoided if the left-right symmetry 
originates from a unified theory in which the $SU(2)_{L,R}$ gaugino
masses are real. As an example, we show how the left-right symmetry can be
embedded in an SO(10) theory.

\vspace{0.4in}


\end{quotation}
\end{titlepage}

\baselineskip=19pt

\section{Introduction}

Quantum Chromodynamics (QCD) is now widely accepted as the theory of strong
interactions. The periodic vacuum structure of QCD has however the unpleasant
implication that strong interactions violate {\it CP}. This 
{\it CP} violating
interaction being flavor conserving only manifests itself as
a electric dipole moment of the neutron and leads to
a value far above the present experimental upper limit unless the associated
{\it CP} violating coupling (usually labelled as $\bar{\Theta}$), which
is left arbitrary by strong interaction dynamics, is somehow suppressed 
to the level of $10^{-9}$. This problem of fine tuning of the $\bar{\Theta}$
parameter in gauge theories is known as the strong {\it CP} 
problem \cite{jkim87}.
There are many solutions to the strong {\it CP} problem\cite{jkim87}: the most
well-known of these is the Peccei-Quinn solution which requires the complete
gauge theory of electroweak and strong interactions to respect a 
global chiral $U(1)$ symmetry. 
This symmetry must however be
spontaneously broken in the process of giving mass to the
W-boson and fermions leading 
to a pseudo-Goldstone boson in the particle spectrum known in
the field as the axion. There are two potential problems with this otherwise
beautiful proposal: (i) the axion has not been experimentally discovered 
as yet and the window is closing in on it; and (ii) if 
non-perturbative gravitational effects induced by black holes and wormholes
are important in particle physics as is believed by some\cite{kami92},
then the axion solution would require fine tuning of the gravitationally
induced couplings by some 50 orders of magnitude. This will make the axion 
theory quite contrived. 

A second class of solution that does not lead to any near massless boson  
is to require the theory to be invariant under discrete 
symmetries\cite{mbeg78,geor78}. In our opinion, the most physically
motivated of such theories are the ones\cite{mbeg78} based on the left-right
symmetric theories of weak interactions\cite{pati74}.
These theories are based on the gauge group
 $SU(2)_L\times SU(2)_R\times U(1)_{B-L}$ 
with quarks and leptons assigned in a left-right symmetric manner.
Such models are also completely quark-lepton symmetric. To see
how parity symmetry of the Lagrangian helps to solve the strong 
{\it CP} problem,
let us note that
the physical QCD induced {\it CP} violating phase can be written as
\be
\bar{\Theta} = \Theta + {\rm Arg} \det (M_u M_d)
\label{eq:theta}
\ee
where $\Theta$ is the parameter in $F \tilde{F}$ part of the QCD
Lagrangian, and $M_u$, $M_d$, are the up and down quark mass
matrices respectively. Invariance under parity sets $\Theta=0$ 
because $F \tilde{F}$ is odd under parity.
Additionally, constraints of left-right symmetry imply that the Yukawa
couplings of quarks responsible for the generation of quark masses
are hermitean. If furthermore the vacuum expectation values (VEVs) of the
Higgs fields responsible are shown to be real, then this would
automatically lead to $\bar{\Theta}=0$ at the tree level. If the one loop
contributions also preserve the hermiticity of the quark mass matrices,
then we have a solution to the strong {\it CP}
problem. In the nonsupersymmetric left-right models with nontrivial
{\it CP} violation, it is well known that in general VEVs of
the Higgs field are not real. This, in the past led to suggestions that
either new discrete symmetries be invoked together with left-right symmetry
or new vectorlike fermions be added to the theory\cite{mbeg78}. 
Such theories also do not suffer from the Planck scale implied 
fine tunings\cite{bere93}. It always
remained a challenge to solve the strong {\it CP} problem only using left-right
symmetry since often new additional symmetries 
invoked are not motivated from any
other consideration.

A second CP related problem is connected with
 the minimal supersymmetric standard
model (MSSM), which is currently a subject of intense discussion
next level of physics beyond the standard model and
is the so called
(usual) SUSY {\it CP} problem \cite{elli82}. 
Namely, in the MSSM the complex phase in the gluino mass is 
arbitrary, and the one-loop gluino contribution to the neutron 
electric dipole moment is larger by two or three orders of 
magnitude than the experimental upper bound.

There have been many proposals in the literature to solve one or
both of these problems. For instance,
one recent suggestion is to consider a
supersymmetric extension of the Peccei-Quinn symmetry \cite{dimo95}
which can solve the strong as well as the SUSY CP problems.
Another proposal in the context of grand unified models 
assumes {\it CP} conserving gaugino masses at 
the GUT scale thereby solving only the SUSY {\it CP} problem \cite{dimo94}.
Other proposals employ spontaneous breaking of {\it CP} symmetry to achieve 
the same goal\cite{babu94}.
None of the above approaches however address the important issue of 
R-parity conservation.

Our goal in this paper is to discuss a possible solution to both the
strong {\it CP} as well as the SUSY {\it CP} problem in supersymmetry. 
The first point to
note is that in supersymmetric theories, the $\bar{\Theta}$ receives an
additional contribution from 
the phase of the gluino mass at the tree level\cite{dine93}:
\begin{eqnarray}
\bar{\Theta}= \Theta + {\rm Arg} \det (M_u M_d) - 3 {\rm Arg} m_{\tilde{g}}
\end{eqnarray}
So any solution to strong {\it CP} problem in supersymmetric theories must also
require that the phase of the gluino mass must be naturally suppressed.
Note that a solution to the SUSY CP problem also requires the suppression of
the same phase though to a lesser degree. Clearly therefore a solution to

In two recent letters\cite{moha96,kuch96}, it has been pointed out that if
supersymmetry is combined with left-right symmetry, the strong {\it CP}
problem is automatically solved without the need for any extra symmetry.
Furthermore, in Ref. \cite{moha96}, it was pointed out that
this model also provides a solution to the SUSY {\it CP} problem
of MSSM, {\it i.e.} it does not lead to large electric dipole moment of the
neutron. As a bonus, these models automatically conserve R-parity.
In this paper we elaborate on the results of Ref.\cite{moha96} and present
some new ones which show the left-right scale independence of our
result. We also discuss the question of possible embedding of left-right
symmetry in grand unified theories.

This paper is organized as follows: in Sec 2, we discuss our
supersymmetric solution to the strong {\it CP} problem; in Sec 3, we
show how the solution remains regardless of whether the right-handed scale 
$M_R$ is in the TeV range or much higher; in Sec 4, we discuss our
solution to the usual SUSY {\it CP} problem;
in Sec 5, we show how
the theory can be embedded into the $SO(10)$ model; in Sec 6, we give our
conclusions. We discuss the question of potential minimization in Appendix
A; show the reality of Higgs VEVs in Appendix B; 
list the evolution equations for Yukawa couplings for
a general four doublet expansion of MSSM in Appendix C;  
and discuss the doublet-doublet splitting in Appendix D. 

\section{Supersymmetric Solution to the Strong {\it CP} Problem}

Let us recall the arguments of Ref. \cite{moha96} and see how
the supersymmetric left-right model solves the strong {\it CP}
problem at the scale $M_R$.

The gauge group of the theory is 
SU$(2)_L \, \times$ SU$(2)_R \, \times$ U$(1)_{B-L}$ with quarks and 
leptons transforming as doublets under SU$(2)_{L,R}$. 
In Table 1, we denote the quark, lepton and Higgs 
superfields in the theory along with their transformation properties
under the gauge group. Note that we have chosen two bidoublet fields
to obtain realistic quark masses and mixings (one bidoublet implies 
a Kobayashi-Maskawa matrix proportional to unity, because supersymmetry forbids
$\tilde{\Phi}$ in the superpotential).

\begin{table}
\begin{tabular}{|c|c|} \hline
Fields           & SU$(2)_L \, \times$ SU$(2)_R \, \times$ U$(1)_{B-L}$ \\
                 & representation \\ \hline
Q                & (2,1,$+ {1 \over 3}$) \\
$Q^c$            & (1,2,$- {1 \over 3}$) \\
L                & (2,1,$- 1$) \\
$L^c$            & (1,2,+ 1) \\
$\Phi_{1,2}$     & (2,2,0) \\
$\Delta$         & (3,1,+ 2) \\
$\bar{\Delta}$   & (3,1,$- 2$) \\
$\Delta^c$       & (1,3,+ 2) \\
$\bar{\Delta}^c$ & (1,3,$- 2$) \\ \hline
\end{tabular}
\caption{Field content of the SUSY LR model}
\end{table}

The superpotential for this theory is given by (we have suppressed
the generation index):

\ba
W & = & 
{\bf Y}^{(i)}_q Q^T \tau_2 \Phi_i \tau_2 Q^c +
{\bf Y}^{(i)}_l L^T \tau_2 \Phi_i \tau_2 L^c 
\nonumber\\
  & +  & i ( {\bf f} L^T \tau_2 \Delta L + {\bf f}_c 
{L^c}^T \tau_2 \Delta^c L^c) 
\nonumber\\
  & +  & \mu_{\Delta} {\rm Tr} ( \Delta \bar{\Delta} ) + 
\mu_{\Delta^c} {\rm Tr} ( \Delta^c \bar{\Delta}^c ) +
\mu_{ij} {\rm Tr} ( \tau_2 \Phi^T_i \tau_2 \Phi_j ) 
\nonumber\\
 & + & W_{\it NR}
\label{eq:superpot}
\ea  
where $W_{\it NR}$ denotes non-renormalizable terms arising from
higher scale physics such as grand unified theories or Planck scale effects.
At this stage all couplings ${\bf Y}^{(i)}_{q,l}$, $\mu_{ij}$,
$\mu_{\Delta}$, $\mu_{\Delta^c}$, ${\bf f}$, ${\bf f}_c$ are 
complex with $\mu_{ij}$, ${\bf f}$ and ${\bf f}_c$ being symmetric matrices. 

The part of the supersymmetric action that arises from this
is given by
\be
{\cal S}_W = \int d^4 x \int d^2 \theta \, W + 
\int d^4 x \int d^2 \bar{\theta} \, W^\dagger \, .
\ee

The terms that break supersymmetry softly to make the theory 
realistic can be written as

\ba
{\cal L}_{\rm soft} & = & \int d^4 \theta \sum_i m^2_i \phi_i^\dagger \phi_i
                      + \int d^2 \theta \, \theta^2 \sum_i A_i W_i 
     + \int d^2 \bar{\theta} \, {\bar{\theta}}^2 \sum_i A_i^* W_i^\dagger
                        \nonumber\\
     & + & \int d^2 \theta \, \theta^2 \sum_p m_{\lambda_p} 
                 {\tilde{W}}_p {\tilde{W}}_p +
           \int d^2 \bar{\theta} \, {\bar{\theta}}^2 \sum_p m_{\lambda_p}^* 
                 {{\tilde{W}}_p}^* {{\tilde{W}}_p}^* \, . 
\label{eq:soft}
\ea

In Eq. \ref{eq:soft},  ${\tilde{W}}_p$ denotes the gauge-covariant
chiral superfield that contains the $F_{\mu\nu}$-type terms with
the subscript going over the gauge groups of the theory including
SU$(3)_c$. $W_i$ denotes the various terms in the superpotential, 
with all superfields replaced by their scalar components and
with coupling matrices which are not identical to those in $W$.
Eq. \ref{eq:soft} gives the most general set of soft breaking terms
for this model.

In Sec. 1 we saw that left-right symmetry implies that the first
term in Eq. \ref{eq:theta} is zero. Let us now see how supersymmetric 
left-right symmetry also requires the second term in this equation to 
vanish naturally. We choose the following definition of left-right 
transformations on the fields and the supersymmetric variable $\theta$

\ba
Q             & \leftrightarrow  &  {Q^c}^* \nonumber\\
L             & \leftrightarrow  &  {L^c}^* \nonumber\\
\Phi_i        & \leftrightarrow  &  {\Phi_i}^\dagger \nonumber\\
\Delta        & \leftrightarrow  &  {\Delta^c}^\dagger \nonumber\\
\bar{\Delta}  & \leftrightarrow  &  {\bar{\Delta}}^{c\dagger} \nonumber\\
\theta        & \leftrightarrow  &  \bar{\theta} \nonumber\\
{\tilde{W}}_{SU(2)_L} & \leftrightarrow  & {\tilde{W}}^*_{SU(2)_R} \nonumber\\  
{\tilde{W}}_{B-L,SU(3)_C} & 
         \leftrightarrow  & {\tilde{W}}^*_{B-L,SU(3)_C}
 \label{eq:lrdef}
\ea 

Note that this corresponds to the usual definition 
$Q_L \leftrightarrow Q_R$, etc. 
With this definition of L-R symmetry, it is easy to check that

\ba
{\bf Y}^{(i)}_{q,l} & = & {{\bf Y}^{(i)}_{q,l}}^\dagger \nonumber\\
\mu_{ij} & = & \mu_{ij}^* \nonumber\\
\mu_\Delta & = & \mu_{\Delta^c}^* \nonumber\\
{\bf f} & = & {\bf f}_c^* \nonumber\\
m_{\lambda_{SU(2)_L}} & = & m_{\lambda_{SU(2)_R}}^* \nonumber\\
m_{\lambda_{B-L,SU(3)_C}} & 
          = & m_{\lambda_{B-L,SU(3)_C}}^*\nonumber\\
A_i & = & A^\dagger_i,
\label{eq:rels}
\ea

We will make extensive use of equations (\ref{eq:rels}) in this 
paper\footnote{Note that the dagger in the last equation for A-terms
indicates that squark mass matrices {\bf h} are hermitean by L-R symmetry,
although they of course do not have to be proportional to Yukawa
mass matrices below some high scale.}. 
The first point to note is that the gluino mass is automatically
real in this model; as a result, the last term in the equation for
$\bar{\Theta}$ above is naturally zero. We now therefore have to
investigate only the quark mass matrices in order to guarantee that
$\bar{\Theta}$ vanishes at the tree level. For this purpose, we note that
the Yukawa matrices are hermitean\footnote{It is interesting that 
more general definitions of left-right
transformations in the flavor sector are possible. For example, invariance
under $Q \to U_1 Q^{c*}$ and $Q^c \to U_2 Q^*$, 
 where $U_1$ and $U_2$ are some $SU(3)$ matrices, 
gives {\it non-}hermitean Yukawa matrices, but which still
have a real determinant.} and
the mass terms involving Higgs bidoublets in the superpotential are real.
If we can show that the vacuum expectation values of the bi-doublets are 
real, then the tree level value of $\bar{\Theta}$ will be naturally zero.

As in \cite{moha96}, for $W_{\rm NR}$ we will use a single operator
$ {\lambda \over M} [ {\rm Tr} ( \Delta^c \tau_m \bar{\Delta}^c) ]^2 $,
in order to be able to have vanishing sneutrino VEVs, as shown in
Appendix A. The $M$ could be equal to $M_{Pl}$ or $M_U$. The other
allowed non-renormalizable operators do not effect our result and 
could be easily included in our discussion.

In this case we have made a detailed 
analysis of the Higgs potential and find that, at the minimum of the 
potential, the $< \Phi_i>$ are always real. This result is not at all 
trivial because of large number of VEVs that enter 
and one might naively think that 
spontaneous {\it CP} violation is possible.  However, a recent analysis
\cite{masi95} has shown that a general supersymmetric model 
with two pairs of Higgs doublets (of which SUSY LR is 
a special case) cannot break {\it CP} spontaneously. We give the 
application of this calculation to the SUSY LR case in Appendix B. 
It is now clear that the quark mass matrices are hermitean and 
therefore $\bar{\Theta} = 0$ naturally at the tree level.

In Ref. \cite{moha96} it was also shown that no strong {\it CP} violating
phase is generated at the one-loop level.
Examples of one-loop diagrams are shown in Figures 1 and 2;
the higgs diagram (Figure 1) and some of the gaugino
diagrams (Figure 2a) generate only hermitean
contributions, while the other gaugino diagrams 
(Figure 2b) is always real, if the gaugino masses are assumed to
be real, as it happens when our model is embedded into a grand unified
theory (see later). Thus in
the total contribution at one-loop level the Yukawa matrices
are still hermitean. We concluded that the first nonzero
contribution to $\bar{\Theta}$, if any, 
arises only at the two loop level, and is thus consistent
with present limits.

\section{Solution to the Strong {\it CP} Problem Holds Below Scale $M_R$}

We have seen how the supersymmetric left-right model solves the
strong {\it CP} problem at the scale of the $SU(2)_R$ breaking $M_R$.
If this scale is of the order of the weak scale then we are done,
because the mass matrices in the expression for $\bar{\Theta}$ are
defined at that scale, and no further phase
 can be generated. Let us investigate what happens if $M_R$ is 
some higher (intermediate) scale.\footnote{Such 
scales can be desired in grand unification schemes with 
LR models as intermediate steps, because of the seesaw scenarios 
of neutrino masses.} Two questions must be answered: 
\begin{itemize}
\item Does the determinant
of the Yukawa matrices stay real below $M_R$? 
\item The one-loop contribution of the
$SU(2)_L$ gaugino is no longer cancelled by the heavy $SU(2)_R$
gaugino. Can we avoid this contribution?
\end{itemize}
As we will show below the answer to both questions is yes.

Above the scale $M_R$ the hermiticity property of Yukawa couplings
stays intact because L-R symmetry is not broken (see Appendix C). 
However, running of the Yukawa matrices below $M_R$ will necesarily spoil
the hermiticity of Yukawa matrices because of breaking of parity
(for example the right handed neutrino is excluded in running
below $M_R$). Thus one might naively think that a nontrivial
$\bar{\Theta}$ will be generated, and that one must put constraints
on $M_R$. However, we will now show that for the simplest case, when the
field content below $M_R$ is that of the MSSM, namely two higgs
doublets, the determinants of the Yukawa matrices stay real.

Let us denote by ${\bf y_i}$ the Yukawa coupling of a Higgs doublet 
$H_i$ (i=1,2). In the MSSM the one-loop running of the Yukawa couplings 
is of the form\cite{mart94}:
\be
{d \over {dt}} {\bf y}_i = {\bf y}_i \, {\bf T}
\label{eq:hrun}
\ee
where ${\bf T}$ is a matrix in flavor space which is a sum 
of terms of the form
${\bf y^\dagger_j y_j}$, ${\rm Tr}({\bf y^\dagger_j y_j}) {\bf 1}$, 
$g^2_a {\bf 1}$ (see Appendix C).
From (\ref{eq:hrun}) one can easily obtain the Jacobi identity
for the determinant
\be
{d \over {dt}} \det {\bf y}_i = \det {\bf y}_i \, {\rm Tr} \, {\bf T}
\label{eq:Jacobi}
\ee
However, ${\rm Tr} \, {\bf T}$ is always real, and since the determinant
of ${\bf y_i}$ is real at the scale $M_R$, it will be real at any
scale below $M_R$. 
We conclude that {\it although the Yukawa matrices will in general
\underline{not} be hermitean anymore at the lower scale, their determinants
will nevertheless stay real}. 

The VEVs of the higgs doublets in MSSM can always be rotated so that both are
real. Thus we conclude that $\bar{\Theta}_{\rm tree} = 0$. 

Let us consider the one-loop contributions to $\bar{\Theta}$ below $M_R$. 
Typical diagrams that contribute at scale $M_R$ are shown in 
Figs. 1 and 2.
Since the running Yukawa matrices have a real determinant the  
diagrams that have at vertices only Yukawa matrices or bidoublet masses
(which are real) will not contribute. 
However, since the right-handed gaugino
will decouple below $M_R$, the phase in the diagram involving the mass of
the left- gaugino will not cancel. The easiest way to circumvent this problem 
is to assume that the gaugino masses are real\cite{kuch96}. It is then easy
to see that in the one-loop running the left gaugino mass stays real.
Indeed, in Section 5 we show that the reality of the $SU(2)_{L,R}$
gaugino masses comes out naturally in an SO(10) model with a 
generalized left-right symmetry.

Let us next address the effect of the trilinear supersymmetry breaking
term involving squarks and the Higgs boson ({\it i.e.} 
${\bf h}_u m_0\tilde{Q}H_u\tilde{u}^c$
and the corresponding term with $u$ replaced by $d$) on $\bar{\Theta}$.
Above the $M_R$ scale, the matrices ${\bf h}_{u,d}$ are hermitean
due to the constraint of left-right symmetry (like the ${\bf Y}_{u,d}$).
Therefore their contribution to $\bar{\Theta}$ involving the gluino
at the one-loop level automatically vanishes above the scale $M_R$.
(Here we used the fact that left-right symmetry requires that the gluino 
masses are real).
As we extrapolate it down to the $M_Z$ scale
using the renormalization group equations\cite{mart94}, we have to
see if the $\det h_{u,d}$ develop any imaginary part. A look at
the one-loop renormalization group equation makes it clear that
such an imaginary part (denoted by $\delta_A$) could develop; let us
therefore estimate its effect on the gluino mass as well as the
quark mass matrices. A rough order of magnitude of the CP violating
phase in the gluino mass can be estimated as follows: since the 
${\bf h}_{u,d}$ are hermitean and proportional to the Yukawa couplings 
${\bf Y}_{u,d}$
at some scale above the $M_R$ scale, let us go to a basis where 
${\bf Y}_d$ and ${\bf h}_d$ are diagonalized. Then we find that, at the scale
of proportionality, if any one of 
the off-diagonal elements of ${\bf Y}_u$ and (hence ${\bf h}_u$) are set to 
zero, the theory becomes
completely CP conserving and cannot generate a CP violating phase 
at any scale below $M_R$. It is then clear that the one loop graph
that generates a phase in the gluino mass can lead to the gluino phase
$\delta_{\tilde{g}}$ which is at most
\ba
\delta_{\tilde{g}}\simeq {{V_{ub}V_{bc}V_{cd}V_{du}\alpha_s}\over{64 \pi^3}}
ln{{M_R}\over{M_Z}}
\ea
leading to $\delta_{\tilde{g}}\leq 10^{-8}$ which is close
to the upper limits on the $\bar{\Theta}$.
Similar arguments can be given for the one loop contribution to
the $\tilde{Q}\tilde{Q}^c$ mass matrix to show that their 
contribution to $\bar{\Theta}$
is around $10^{-8}$ .

It is worth pointing out at this stage that in the above discussion
we have assumed that the theory below $M_R$ is the MSSM (except of course
the fact that the ``obnoxious" R-parity violating terms are 
naturally absent). In Appendix D, we discuss one way of obtaining
MSSM in the framework of our model.

\vskip 0.6cm

At the end, let us consider what happens if we consider an effective
four higgs doublet model below $M_R$. The running of Yukawa couplings at
one loop are listed in Appendix C. We note that the running of 
Yukawa matrices does not have a form of (\ref{eq:hrun}). There are
additional terms on the right hand side of the form 
${\bf y}_j {\rm Tr} ({\bf y}^\dagger_j {\bf y}_i)$ ($i \neq j$), 
thus invalidating the
Jacobi identity for determinants. Indeed, such a term will in general
produce phases of order $ V_{cb}^2 / (16 \pi^2) \approx 10^{-5}-10^{-6}$.
In this case, we also expect additional suppression coming from the fact
at some very high scale, the theory becomes CP conserving
if any off-diagonal element of $Y_u$ is set to zero.
Barring enough suppress from this, it may be necessary to
impose some additional symmetry to
suppress the Yukawa couplings of the second  pair of Higgs 
doublets\cite{masi95}. \footnote{ Note however
that in the second paper of Ref. \cite{masi95}, a general four higgs doublet 
model with arbitrary Yukawa couplings was considered, and the
additional symmetry was needed to suppress too large {\it CP} violation
in $K\bar{K}$ mixing; strong {\it CP} violation was too large in that model.
In our case Yukawa couplings have the constraint that they
come from hermitian matrices at the $M_R$ scale, and the additional
global symmetry is enough to solve the strong {\it CP} problem.}

In conclusion, if the effective theory below
$M_R$ has the MSSM-like field content and if the left gaugino mass
is real, no observable $\bar{\Theta}$ will be generated for all
values of $M_R$ from some intermediate scale ($\approx 10^{12}$ GeV)
all the way down to 1 TeV.

\vskip 0.5cm

\section{Solution to the SUSY {\it CP} Problem}

\vskip 0.5cm

Let us now turn to the discussion of the SUSY CP problem. The main issue
here is the potentially large contribution to the electric dipole
moment of the neutron at the one-loop level. An analysis of the
various aspects of the problem has been reviewed in Ref.\cite{gari93}.
In the standard parameterization
of the MSSM interactions at the electroweak scale, the large contributions
to $d^n_e$ comes from two sources: the phases of the $(Am_{\tilde{g}})$
and $(\mu v_u m_{\tilde{g}}/ v_d)$ terms. Another way to state this
is to note that the first term 
originate from the same trilinear scalar
susy breaking terms $h_{u,d}$ discussed in the previous section,
whereas the second term arises from the F-term contribution extrapolated
down to the electroweak scale. We work in a basis where the diagonal
block matrices in the squark ($\tilde{q}-\tilde{q}^c$) mass matrices are
diagonalized. We will then be interested in the $11$ entry of the gluino
one loop contribution to electric dipole moment operator for both the up
and the down sector.
 
First point to note is that in our model, above the $M_R$ scale, the
hermiticity of ${\bf h}_{u,d}$ and ${\bf Y}_{u,d}$ together with the
reality of the gluino mass implies that  
 there is no one-loop contribution to $d^n_e$. Garisto\cite{gari93}
has argued that if the above parameters are real at any high energy
scale, their contribution to $d^n_e$ remains small at the electroweak
scale. For example, in our case as shown above, the phase of the gluino 
mass at the scale $M_Z$ is of order $10^{-8}$ . As far as the
the ${\bf h}_{u,d}$ and ${\bf Y}_{u,d}$ terms are concerned, we have not
succeeded in showing that once extrapolated down to the $M_Z$ scale, 
the $11$ term of the gluino induced dipole moment matrix
remains real. However, using the already stated argument above,
the hermiticity of the Yukawa matrices above the scale $M_R$ implies that
 that any departure from reality is at most of order
$\delta\equiv  V_{ub}V_{bc}V_{cd}/(16\pi^2)ln{{M_R}\over{M_Z}}\simeq 
10^{-8}$. 
We then expect that the maximum contribution to $d^n_e$ from them to
\be
(d^n_e)^{max} \leq { {8e\alpha_s m_d} \over {27\pi M^2_{\tilde{q}}}}\delta
 \leq 4\times 10^{-31}~~ecm
\label{dipole} 
\ee 
where we have assumed that $m_d\simeq 10$ MeV and $M_{\tilde{q}}\simeq 100$
GeV. This is safely within the present experimental upper limit.
Thus our model provides simultaneously a solution to the SUSY CP
problem without the need for any new symmetries.

Let us note that in this paper we do not address the usual SUSY 
flavor problems with $K\bar{K}$ mixings,
etc., that require a certain degree of fine tuning in the structure 
of squark and quark matrices, since this is beyond the scope of
our paper. Left-right symmetry implies that the Yukawa matrices
be hermitean, but the flavor properties such as hierarchy and
alignment must come from the underlying flavor theory, and
we  refer the reader to the existing solutions \cite{ynir93}
which may be employed here as well.

\vskip 0.5cm

\section{$SO(10)$ Embedding and Reality of Weak Gaugino Masses}

\vskip 0.5cm

In this Section, we address the question of embedding the left-right
model into an $SO(10)$ theory so that we not only have a grand unified version
of our theory but also we guarantee the reality of the gaugino masses
(i.e.$m_{\lambda_{L,R}}= m^*_{\lambda_{L,R}}$). The reality of the gaugino
masses follow from the combination of two things: the requirement of left-right
symmetry implies as shown earlier that $m_{\lambda_L}=m^*_{\lambda_R}$;
on the other hand $SO(10)$ unification implies that the two gauginos being
part of the same {\bf 45} dimensional representation have equal mass. The
main task for us in this section is to show that there exists a definition
of left-right symmetry which preserves the hermiticity of the Yukawa couplings.

To prove the hermiticity of the Yukawa couplings, we will exhibit
only the simplest model and not attempt to
address the issues such as doublet triplet splitting etc. 
Let us consider the Higgs fields belonging to
{\bf 10}(denoted by {\bf H}), {\bf 45}(denoted {\bf A}) and {\bf 126}
(denoted by $\Delta$)(plus ${\bf {\bar{\Delta}}}$) representations. The
superpotential involving all these fields can be written as:
\ba
W_{GUT} & = & h^{s}_{ab}\psi^T_a B \Gamma_i\psi H_i +
h^{A}_{ab}\psi^T_a B\Gamma_i\Gamma_j\Gamma_k\psi_b(H_iA_{jk})_{(Anti)}/{M}
\nonumber\\
 & + & h'_{ab}\psi^T_a B \Gamma_i \psi_b H_jA_{ij}/M
+f_{ab} \psi^T_a B \Gamma_i\Gamma_j\Gamma_k\Gamma_l\Gamma_m \psi_b 
\Delta_{ijklm} 
\nonumber\\
& + & {\rm terms} \, {\rm involving} \, \Delta \, 
{\rm in} \, {\rm order} \, {1 \over M}
\label{SO10}
\ea
It is well known that $h^{s}$, $h'$ and $f$ are symmetric matrices whereas
$h^{A}$ is an antisymmetric combination since we have projected out the
{\bf 120} dim. representation from the {\bf 10} and {\bf 45} product in
the $h^{A}$ term. Let us now define that under parity transformation
\be 
\psi\rightarrow D_{\psi}B^{-1}\psi^* \, , \, 
H_i\rightarrow -D_{H}H^*_iD^{-1}_{H} \, , \,
A_{ij}\rightarrow -D_{A} A^*_{ij}D^{-1}_{A} \, , \,
\Delta \rightarrow -D_{\Delta}\Delta^*D^{-1}_{\Delta}
\label{trans}
\ee
Here $B$ is the charge conjugation matrix for $SO(10)$; $D$ is the
operator that implements the left-right transformation inside the
$SO(10)$ multiplets $\psi$, $H$ etc.\footnote{This is similar to 
the L-R definition (\ref{eq:lrdef}). This comes because, for example,
strictly speaking $Q^c$ is not a doublet under $SU(2)_R$, but rather
$Q^{'c} \equiv \tau_2 Q^c$. So the L-R definition (\ref{eq:lrdef}) in
terms of the gauge multiplets would be: $Q \leftrightarrow \tau_2 Q^{'c*}$.
The operator $\tau_2$ plays a similar role as the operator $D$ above
in the $SO(10)$ case.}
For instance operating on the {\bf 10}
dimensional representation ($H_i$), it changes $H_7\rightarrow -H_7$ and
leaves all other components unchanged; Similarly in {\bf 45}, it changes
the sign of all elements that carry the index $7$  etc. (Note that the
choice of the 7th component is basis dependent and we are working in
a basis where $I_{3L}={{1}\over{4}}(\Sigma_{90}-\Sigma_{78})$ and
$I_{3R}={{1}\over{4}}(\Sigma_{90}+\Sigma_{78})$, where $\Sigma_{ij}$
are the generators of $SO(10)$.)

Now, using the fact that $B^T=-B$ and $B^{-1}\Gamma_i B= -\Gamma^T_i$,
it is then easy to show that $h^s_{ab}$, $h'_{ab}$ and $f_{ab}$ are real
whereas $h^{A}_{ab}$ is imaginary. Together with the symmetricity properties
this implies that all this matrices are hermitean.

Now if the $SO(10)$ symmetry is
broken down to $SU(2)_L\times SU(2)_R\times U(1)_{B-L}\times SU(3)_c$
by the {\bf 45} VEV as $\langle A \rangle=i\tau_2 diag( v,v,v,0,0)$,
then the effective low energy theory has two bi-doublets and also
general {\it hermitean} Yukawa coupling of quarks. This leads to the 
embedding of our solution to the strong CP problem in an $SO(10)$ model.

We do not discuss here the unification of gauge couplings in 
theories with $SU(2)_L \times SU(2)_R$ as intermiediate symmetries, 
but note that examples of successful scenarios exist which could
implement our mechanism \cite{bena94}.

\vskip 0.5cm

\section{Conclusion}

\vskip 0.5cm

We have shown that if the minimal supersymmetric extension
of the standard model (MSSM) is embedded in 
the supersymmetric left-right model at higher energies, both the strong
and weak CP problems of the MSSM are automatically cured. Adding this
to the already known result that the R-parity conservation is restored
 as an exact symmetry in the SUSYLR model, thereby providing a naturally stable
neutralino that can act as the cold dark matter of the universe, makes
this embedding quite attractive. The left-right symmetry is then
incorporated into an $SO(10)$ grand unified theory where the scale
of right handed symmetry breaking may be quite high. We show how the
conclusion about the vanishing of the strong CP parameter remains
unchanged in this case.

\vskip 0.3cm

{\bf Acknowledgments}

\vskip 0.3cm

A.R. thanks Markus Luty and Jogesh Pati for useful discussions. 
R.N.M. would like to
thank Alex Pomarol for several important comments and discussions.
This work was supported by the NSF grant No. PHY $9421385$.
The work of R.N.M. is also partially suported by the Distinguished
Faculty Research award by the University of Maryland. R.N.M. would
like to acknowledge the hospitality of the CERN Theory Division during
the last part of the work.

\vskip 0.5cm

\vskip 0.6cm

\noindent{\bf APPENDIX A:  Avoiding Sneutrino VEVs}

\vskip 0.6cm

In this Appendix we will show that if in the minimal SUSY LR model one 
includes non-renormalizable Planck scale induced terms, the ground state
of the theory can be $Q^{em}$ conserving even for 
$<{\tilde{\nu}}^c> = 0$. For this purpose, let us briefly recall the argument
of Ref. \cite{kuch93}. The part of the potential containing 
${\tilde{l}}^c$, $\Delta^c$ and ${\bar{\Delta}}^c$ fields only has 
the form (see Appendix B or \cite{kuch93} )

\be
V = V_0 + V_D \, ,
\ee

where

\ba
V_0 & = &
{\rm Tr} | i {\bf f}^\dagger L^c {L^c}^T \tau_2 + 
\mu_\Delta^* \bar{\Delta}^c|^2 \nonumber\\
& + &
\mu^2_1 {\rm Tr} ( \Delta^c \Delta^{c\dagger}) \, +
\mu^2_2 {\rm Tr} ( {\bar{\Delta}}^c {\bar{\Delta}}^{c\dagger} ) \nonumber\\
& + &
\mu^2_3 {\rm Tr} (\Delta^c \bar{\Delta}^c ) +
\mu_4 {\tilde{L}}^{cT} \tau_2 \Delta^c L^c \, ,  
\ea 

and

\ba
V_D & = & 
{g^2 \over 8} \sum_{m} | {\tilde{L}}^{c\dagger} \tau_m {\tilde{L}}^c +
{\rm Tr} ( 2 {\Delta}^{c\dagger} \tau_m {\Delta}^c + 
     2 {\bar{\Delta}}^{c\dagger} \tau_m {\bar{\Delta}}^c  |^2 \nonumber\\
   & + &
{g'^2 \over 8} | {\tilde{L}}^{c\dagger} {\tilde{L}}^c -
2 \, {\rm Tr} ( {\Delta}^{c\dagger} {\Delta}^c -
{\bar{\Delta}}^{c\dagger} {\bar{\Delta}}^c ) |^2 \, .
\ea

Note that if $<{\tilde{\nu}}^c> = 0$ then the vacuum state for which
$\Delta^c = { 1 \over \sqrt{2} } v \tau_1 $ and
${\bar{\Delta}}^c = { 1 \over \sqrt{2} } v' \tau_1 $ is 
lower than the vacuum state
$
\Delta^c = v
\left (
\begin{array}{cc}
0 &  0 \\ 
1 &  0 \\
\end{array} \right ) $ and
$
{\bar{\Delta}}^c = v'
\left (
\begin{array}{cc}
0 &  1 \\ 
0 &  0 \\
\end{array} \right ) \, .
$
However, the former is electric charge violating. The only way to 
have the global minimum conserve electric charge is to have
$<{\tilde{\nu}}^c> \neq 0$. On the other hand, if we have 
non-renormalizable terms included in the theory, the situation changes:
for instance, let us include non-renormalizable gauge invariant terms of the
form (inclusion of other non-renormalizable terms simply enlarges the
parameter space where our conclusion holds):

\be
W_{NR} = {\lambda \over M} 
[ {\rm Tr} ( \Delta^c \tau_m {\bar{\Delta}}^c) ]^2 \, .
\ee
This will change V to the form:

\be
V = V_0 + V_{NR} + V_D \, ,
\ee 
where $V_0$ and $V_D$ are given before and $V_1$ is given by
\be
V_{NR} = {{\lambda \mu} \over M} 
[ {\rm Tr} ( \Delta^c \tau_m {\bar{\Delta}}^c )]^2
+ {{4 \lambda \mu_\Delta} \over M} 
[ {\rm Tr} ( \Delta^c \tau_m {\bar{\Delta}}^c )]
[ {\rm Tr} ( \Delta^{c\dagger} \tau_m \Delta^c ) ]
+ \Delta^c \leftrightarrow {\bar{\Delta}}^c + {\rm etc.} 
\ee
For the charge violating minimum above, this term vanishes but the 
charge conserving minimum receives a nonzero contribution. 
Note that the sign of $\lambda$ is arbitrary and therefore, by 
appropriately choosing  sgn($\lambda$) we can make the
electric charge conserving vacuum lower than the $Q^{em}$-violating one.
In fact, one can argue that, since we expect
$v^2 - v'^2 \approx { {f^2 (M_{SUSY})^2} \over {16\pi^2} }$ in typical 
Polonyi type models, the charge conserving minimum occurs for
$ f < 4 \pi 
\left( { {4 \lambda \mu_\Delta} \over M } \right)^{1 \over 4}
{{v}\over{M_{SUSY}}}$.
For $\lambda \approx 1$, $\mu_\Delta \approx v \approx M_{SUSY} 
\approx 1 {\rm TeV}$ and $M= M_{Pl}$,
we get $f \leq 10^{-3}$ if $v-M_{SUSY}$. 
We have assumed that the right handed scale
is in the TeV range.
 The constraint on $f$ of course becomes weaker for 
larger values of $\mu_\Delta$. We wish to note that a possible 
non-renormalizable term of the form 
$  \lambda_1 {\rm Tr} ( \Delta^c \tau_m {\bar{\Delta}}^c ) 
{\rm Tr} ( \Phi_i \tau_m \Phi_j ) { 1 \over M_{Pl} } $
does induce a complex effective mass for the bidoublets but its magnitude 
is given by ${ v_R^2 \over M_{Pl} }$ which for 
$v_R \leq 10^{10} / \lambda_1$ GeV gives a phase 
of order $10^{-9}$ and is negligible if $v_R$ is in the TeV range.
Its presence therefore does not affect the solution to 
the strong {\it CP} problem outlined in the paper for all values of
$v_R$ from TeV up to some intermediate scale $\approx 10^{11}-10^{12}$ GeV,
depending on the value of $\lambda_1$ . 

Furthermore, it is also important to point out that 
since Planck scale effects
are not expected to respect any global symmetries, the coupling
parameters of the higher dimensional
terms in Eq. (16) involving $\Delta$ and $\Delta^c$ will be different.
This difference will help in the realization of the parity violating
minimum as the global minimum of the theory.

\vskip 0.3cm

\noindent{\bf APPENDIX B: Reality of Bidoublet VEVs}

\vskip 0.6cm

Here we show that the VEVs of the bidoublet Higgs fields in the 
supersymmetric left-right model are real. The scalar potential is given by

\be
V = V_F + V_{\rm soft} + V_D + V_{NR}(\Delta^c, \bar{\Delta}^c)\, ,
\ee
where

\ba
V_F & = &\sum_{p} | {{\bf Y}^{(i)}_q}_{pr} \tau_2 \Phi_i \tau_2 Q^c_r|^2 +
\sum_{r} | {{\bf Y}^{(i)}_q}_{pr} Q_p \tau_2 \Phi_i \tau_2 |^2 \nonumber\\ 
& + &
\sum_{i} {\rm Tr} | {\bf Y}^{(i)T}_q Q Q^{cT} + {\bf Y}^{(i)T}_l L L^{cT} + 
                     2 \mu_{ij} \Phi_j|^2 \nonumber\\   
& + &
\sum_{p} | {{\bf Y}^{(i)}_l}_{pr} \tau_2 \Phi_i \tau_2 L^c_r
+ 2 i f_{pr} \tau_2 \Delta L_r |^2 +
\sum_{r} | {{\bf Y}^{(i)}_l}_{pr} L_p \tau_2 \Phi_i \tau_2
+ 2 i f_{pr}^* L^c_p \tau_2 \Delta^c |^2  \nonumber\\
& + &
{\rm Tr} | i {\bf f}^T L L^T \tau_2 + \mu_\Delta \bar{\Delta}|^2 +
{\rm Tr} | i {\bf f}^\dagger L^c {L^c}^T \tau_2 + 
\mu_\Delta^* \bar{\Delta}^c|^2 \nonumber\\
& + &
|\mu_\Delta|^2 {\rm Tr} 
( \Delta \Delta^\dagger + \Delta^c \Delta^{c\dagger}) \, ,
\ea 

\ba
V_{\rm soft} & = & m^2_q ({\tilde{Q}}^\dagger \tilde{Q} + 
                           {\tilde{Q}}^{c\dagger} {\tilde{Q}}^c)
  + m^2_l ({\tilde{L}}^\dagger \tilde{L} + 
                           {\tilde{L}}^{c\dagger} {\tilde{L}}^c)
  + m^2_{\Phi_i} \Phi_i^\dagger \Phi_i
\nonumber\\
  & + & 
m^2_{\Delta} {\rm Tr} ( \Delta^\dagger \Delta + {\Delta^c}^\dagger \Delta^c )
+  m^2_{\bar{\Delta}} 
 {\rm Tr} ( {\bar{\Delta}}^\dagger \bar{\Delta} 
   + {\bar{\Delta}}^{c\dagger} {\bar{\Delta}}^c ) \nonumber\\
  & + &
[ \, 
A_{q,i} {\bf Y}^{(i)}_q {\tilde{Q}}^T \tau_2 \Phi_i \tau_2 {\tilde{Q}}^c +
A_{l,i} {\bf Y}^{(i)}_l {\tilde{L}}^T \tau_2 \Phi_i \tau_2 {\tilde{L}}^c 
\nonumber\\
  & +  & A_L i ( {\bf f} {\tilde{L}}^T \tau_2 \Delta L 
          + {\bf f}^* {\tilde{L}}^{cT} \tau_2 \Delta^c L^c) 
\nonumber\\
  & +  & A_\Delta ( \mu_{\Delta} {\rm Tr} (\Delta \bar{\Delta}) + 
              \mu_{\Delta}^* {\rm Tr} (\Delta^c \bar{\Delta}^c ) ) +
A_\Phi \mu_{ij} {\rm Tr} ( \tau_2 \Phi^T_i \tau_2 \Phi_j ) 
+ {\rm h.c.} ] \, ,
\ea 

\ba
V_D & = & 
{g^2 \over 8} \sum_{m} | {\tilde{L}}^\dagger \tau_m \tilde{L} +
{\rm Tr} ( 2 \Delta^\dagger \tau_m \Delta + 
     2 {\bar{\Delta}}^\dagger \tau_m \bar{\Delta} +
     \Phi^\dagger \tau_m \Phi ) |^2 \nonumber\\
   & + & 
{g^2 \over 8} \sum_{m} | {\tilde{L}}^{c\dagger} \tau_m {\tilde{L}}^c +
{\rm Tr} ( 2 {\Delta}^{c\dagger} \tau_m {\Delta}^c + 
     2 {\bar{\Delta}}^{c\dagger} \tau_m {\bar{\Delta}}^c +
     \Phi \tau_m^T \Phi^\dagger ) |^2 \nonumber\\
   & + &
{g'^2 \over 8} | {\tilde{L}}^{c\dagger} {\tilde{L}}^c -
{\tilde{L}}^\dagger \tilde{L} +
2 \, {\rm Tr} ( \Delta^\dagger \Delta -
{\Delta}^{c\dagger} {\Delta}^c -
{\bar{\Delta}}^\dagger \bar{\Delta} +
{\bar{\Delta}}^{c\dagger} {\bar{\Delta}}^c |^2 \, ,
\ea
and $V_{NR}$ is defined in Appendix A.

We assume the following fields get the VEVs:
\be
< \Delta^c >=
\left (
\begin{array}{cc}
0 &  0 \\ 
\Delta^0 e^{-i\beta_\Delta} &  0 \\
\end{array} \right ) \, ,
< {\bar{\Delta}}^c >=
\left (
\begin{array}{cc}
0 &  \delta^0 \\ 
0 &  0 \\
\end{array} \right ) \, ,
\ee
and
\be
< \Phi_1 >=
\left (
\begin{array}{cc}
v_1 &  0 \\ 
0 &  v_2 e^{i\delta_2} \\
\end{array} \right ) \, ,
< \Phi_2 > =
\left (
\begin{array}{cc}
v_3 e^{i\delta_3} &  0 \\ 
0 &  v_4 e^{i\delta_4} \\
\end{array} \right ) \, ,
\ee
where we have rotated away the nonphysical phases.

The VEV of the scalar potential is
\ba
<V> & = &
|2 \mu_{11} v_1 + 2 \mu_{12} v_3 e^{i\delta_3}|^2 +
|2 \mu_{11} v_2 e^{i\delta_2} + 2 \mu_{12} v_4 e^{i\delta_4}|^2 \nonumber\\
    & + & 
|2 \mu_{21} v_1 + 2 \mu_{22} v_3 e^{i\delta_3}|^2 +
|2 \mu_{21} v_2 e^{i\delta_2} + 2 \mu_{22} v_4 e^{i\delta_4}|^2 \nonumber\\
    & + &
|\mu_\Delta|^2 ( \Delta^{0 \, 2} + \delta^{0 \, 2} ) \nonumber\\
    & + &  m^2_\Delta \Delta^{0 \, 2} 
+ m^2_{\bar{\Delta}} \delta^{0 \, 2} \nonumber\\ 
    & + & 
m^2_{\Phi_1} (v^2_1 + v^2_2) +
m^2_{\Phi_2} (v^2_3 + v^2_4) +
A_\Delta |\mu_\Delta| \Delta^0 \delta^0 
\cos ( \beta_\Delta + {\rm Arg}(\mu_\Delta)) \nonumber\\
    & + &
A_\Phi \mu_{11} 2 v_1 v_2 \cos \delta_2 +
A_\Phi \mu_{12} 2 ( v_1 v_4 \cos \delta_4 + 
v_2 v_3 \cos ( \delta_2 + \delta_3 ) ) \nonumber\\
    & + &
A_\Phi \mu_{22} 2 v_3 v_4 \cos (\delta_3 + \delta_4)
   + <V_D> + V_{NR}(< \Delta^c >, < \bar{\Delta}^c >)
\ea
where $<V_D>$ is the VEV of the D-term

\ba
<V_D> & = & {g^2 \over 8} [ v^2_1 + v^2_3 - v^2_2 - v^2_4 ]^2 \nonumber\\
      & + & {g^2 \over 8} 
[ 2 ( \Delta^{0 \, 2} + \delta^{0 \, 2} )
+ v^2_1 + v^2_3 - v^2_2 - v^2_4 ]^2 \nonumber\\
      & + & {{g'^2} \over 8} 
[ 2 ( \Delta^{0 \, 2} + \delta^{0 \, 2} ) ]^2 \, .
\ea

Note that the phases of the bidoublets $\delta_i$, $i = 2,3,4$
come in the following terms
\ba 
&& v_1 v_i  \cos \delta_i \, , \, i=2,3,4 \nonumber\\
&& v_2 v_3  \cos (\delta_2 + \delta_3 )\nonumber\\
&& v_2 v_4  \cos (\delta_2 - \delta_4 )\nonumber\\
&& v_3 v_4  \cos (\delta_3 + \delta_4 )
\ea
Also, powers of the bidoublet VEVs which are higher than two come only in
the D-term, and there in one only combination 
$g({\bf v}) = v^2_1 + v^2_3 - v^2_2 - v^2_4$. This is exactly the 
situation in general four Higgs doublet supersymmetric models with 
real mass parameters. In Ref. \cite{masi95} it was shown, by using 
a simple geometrical interpretation for the minimum equations for 
the three phases, that the minimum in such a model is {\it CP} conserving.
Thus we conclude that in the SUSY LR model the VEVs of the doublets are real. 
This conclusion holds for general $A_{\Phi_{ij}}$, which can be different 
for different i,j.

The phase of the VEV of the triplet $\beta_\Delta$ is in general non-zero
({\it e.g.} induced by the phase of the coupling $\mu_\Delta$) 
but it does not couple 
to the VEVs of the doublets. Thus it is irrelevant since it does not
enter the calculation of $\bar{\Theta}$ at the tree level or
one loop.

\vskip 0.3cm

\noindent{\bf APPENDIX C: One-loop Running of Yukawa 
Couplings}

\vskip 0.6cm

{\bf a) Four Higgs Doublet SUSY Model}

Here we list the one-loop running of Yukawa couplings for a general {\it four}
higgs doublet supersymmetric model. The Yukawa matrices of 
Higgses that couple to down quarks are denoted by ${\bf y_1}$ and 
${\bf y_3}$, and similarly ${\bf y_2}$ and ${\bf y_4}$ for the up type 
quarks:
\be
L_Y = Q {\bf y_1} D H_1 + Q {\bf y_3} D H_3 + 
Q {\bf y_2} U H_2 + Q {\bf y_4} U H_4 +
Q {\bf y^e_1} E H_1 + Q {\bf y^e_3} E H_3
\ee

\ba
{d \over {dt}} {\bf y_1}  & =  & { 1 \over {16 \pi^2}} \{ {\bf y_1}
[ {\rm Tr} (3 {\bf y_1^\dagger} {\bf y_1}  + {\bf y_1^{e\dagger}} {\bf y^e_1} )
+ 3 {\bf y_1^\dagger} {\bf y_1} + {\bf y_2^\dagger} {\bf y_2}
+   {\bf y_3^\dagger} {\bf y_3} +   {\bf y_4^\dagger} {\bf y_4} ]\nonumber\\
& + & {\bf y_3} [ {\rm Tr} (3 {\bf y_3^\dagger} {\bf y_1}  
+ {\bf y_3^{e\dagger}} {\bf y^e_1} ) + 2 {\bf y_3^\dagger} {\bf y_1} ] 
- {\bf y_1} ( {7 \over 15} g_1^2 + 3 g_2^2 + {16 \over 3} g_3^2) \}\nonumber\\
{d \over {dt}} {\bf y_2}  & =  & { 1 \over {16 \pi^2}} \{ {\bf y_2}
[ {\rm Tr} (3 {\bf y_2^\dagger} {\bf y_2} )
+  {\bf y_1^\dagger} {\bf y_1} + 3 {\bf y_2^\dagger} {\bf y_2}
+   {\bf y_3^\dagger} {\bf y_3} +   {\bf y_4^\dagger} {\bf y_4} ]\nonumber\\
& + & {\bf y_4} [ {\rm Tr} (3 {\bf y_4^\dagger} {\bf y_2}) 
+ 2 {\bf y_4^\dagger} {\bf y_2} ]
- {\bf y_2} ( {13 \over 15} g_1^2 + 3 g_2^2 + {16 \over 3} g_3^2) \}\nonumber\\
{d \over {dt}} {\bf y_3}  & =  & { 1 \over {16 \pi^2}} \{ {\bf y_3}
[ {\rm Tr} (3 {\bf y_3^\dagger} {\bf y_3}  + {\bf y_3^{e\dagger}} {\bf y^e_3} )
+ {\bf y_1^\dagger} {\bf y_1} + {\bf y_2^\dagger} {\bf y_2}
+  3 {\bf y_3^\dagger} {\bf y_3} +   {\bf y_4^\dagger} {\bf y_4} ]\nonumber\\
& + & {\bf y_1} [ {\rm Tr} (3 {\bf y_1^\dagger} {\bf y_3}  
+ {\bf y_1^{e\dagger}}{\bf y^e_3} ) + 2 {\bf y_1^\dagger} {\bf y_3} ]
- {\bf y_3} ( {7 \over 15} g_1^2 + 3 g_2^2 + {16 \over 3} g_3^2) \}\nonumber\\
{d \over {dt}} {\bf y_4}  & =  & { 1 \over {16 \pi^2}} \{ {\bf y_4}
[ {\rm Tr} (3 {\bf y_4^\dagger} {\bf y_4} )
+  {\bf y_1^\dagger} {\bf y_1} +  {\bf y_2^\dagger} {\bf y_2}
+   {\bf y_3^\dagger} {\bf y_3} + 3 {\bf y_4^\dagger} {\bf y_4} ]\nonumber\\
& + & {\bf y_2} [ {\rm Tr} (3 {\bf y_2^\dagger} {\bf y_4})
+ 2 {\bf y_2^\dagger} {\bf y_4}) ] 
- {\bf y_1} ( {13 \over 15} g_1^2 + 3 g_2^2 + {16 \over 3} g_3^2) \}\nonumber\\
{d \over {dt}} {\bf y^e_1}  & =  & { 1 \over {16 \pi^2}} \{ {\bf y^e_1}
[ {\rm Tr} (3 {\bf y_1^\dagger} {\bf y_1}  + {\bf y_1^{e\dagger}} {\bf y^e_1} )
+ 3 {\bf y_1^{e\dagger}} {\bf y^e_1} + 
   {\bf y_3^{e\dagger}} {\bf y^e_3} ]\nonumber\\
& + & {\bf y^e_3} [ {\rm Tr} (3 {\bf y_3^\dagger} {\bf y_1}  
+ {\bf y_3^{e\dagger}} {\bf y^e_1} ) + 2 {\bf y_3^{e\dagger}} {\bf y^e_1} ]
- {\bf y^e_1} ( {9 \over 5} g_1^2 + 3 g_2^2 ) \}\nonumber\\
{d \over {dt}} {\bf y^e_3}  & =  & { 1 \over {16 \pi^2}} \{ {\bf y^e_3}
[ {\rm Tr} (3 {\bf y_3^\dagger} {\bf y_3}  + {\bf y_3^{e\dagger}} {\bf y^e_3} )
+  {\bf y_1^{e\dagger}} {\bf y^e_1} + 
  3 {\bf y_3^{e\dagger}} {\bf y^e_3} ]\nonumber\\
& + & {\bf y^e_1} [ {\rm Tr} (3 {\bf y_1^\dagger} {\bf y_3}  
+ {\bf y_1^{e\dagger}} {\bf y^e_3} ) + 2 {\bf y_1^{e\dagger}} {\bf y^e_3} ]
- {\bf y^e_3} ( {9 \over 5} g_1^2 + 3 g_2^2 ) \} 
\label{eq:fourh}
\ea

The equations for a {\it two} higgs doublet model (i.e. the MSSM) are easily 
obtained by setting for example Yukawa matrices ${\bf y_3}$, ${\bf y_4}$ 
and ${\bf y^e_3}$ to zero in the equations above. They indeed have the
form of Eq. \ref{eq:hrun}.

We see that the equations (\ref{eq:fourh}) part from the form of 
(\ref{eq:hrun}) because of higgs doublet wave function renormalization 
terms. (For example the term 
${\bf y_3} {\rm Tr} (3 {\bf y_3^\dagger} {\bf y_1})$ 
in the equation for ${\bf y_1}$.) One can still write an equation
in form of (\ref{eq:Jacobi}) with new terms in ${\bf T}$ which are
not real in general. For example the phase will appear in 
${\bf y_1^{-1}} {\bf y_3} {\rm Tr} (3 {\bf y_3^\dagger} {\bf y_1})$,
and it will depend on the structure of the Yukawa matrices how big the phase
is.

{\bf b) SUSY LR Model}

It is easy to generalize the above one-loop runnings for the case
of Yukawa couplings in the SUSY LR model with two bidoublets:
\be
L_Y = Q {\bf Y_1} Q^c \Phi_1 + Q {\bf Y_2} Q^c \Phi_2 + 
L {\bf Y^e_1} L^c \Phi_1 + L {\bf Y^e_2} L^c \Phi_2 
\ee
We simply take ${\bf y_1} = {\bf y_2} = {\bf Y_1}$ (and similarly for other
Yukawas), add the right-handed neutrino and compute the contribution
from gauge couplings. Alternatively, we use general formulas \cite{mart94}.
In any case we obtain: 

\ba
{d \over {dt}} {\bf Y_1}  & =  & { 1 \over {16 \pi^2}} \{ {\bf Y_1}
[ {\rm Tr} (3 {\bf Y_1^\dagger} {\bf Y_1}  + {\bf Y_1^{e\dagger}} {\bf Y^e_1} )
+ 4 {\bf Y_1^\dagger} {\bf Y_1} + 2 {\bf Y_2^\dagger} {\bf Y_2} ]\nonumber\\
& + & {\bf Y_2} [ {\rm Tr} (3 {\bf Y_2^\dagger} {\bf Y_1}  
+ {\bf Y_2^{e\dagger}} {\bf Y^e_1} ) + 2 {\bf Y_2^\dagger} {\bf Y_1} ] 
- {\bf Y_1} ( {4 \over 9} g_{B-L}^2 + 3 g_L^2 + 3 g_R^2
{16 \over 3} g_3^2) \}\nonumber\\
{d \over {dt}} {\bf Y_2}  & =  & { 1 \over {16 \pi^2}} \{ {\bf Y_2}
[ {\rm Tr} (3 {\bf Y_2^\dagger} {\bf Y_2}  + {\bf Y_2^{e\dagger}} {\bf Y^e_2} )
+ 2 {\bf Y_1^\dagger} {\bf Y_1} + 4 {\bf Y_2^\dagger} {\bf Y_2} ]\nonumber\\
& + & {\bf Y_1} [ {\rm Tr} (3 {\bf Y_1^\dagger} {\bf Y_2}  
+ {\bf Y_1^{e\dagger}} {\bf Y^e_2} ) + 2 {\bf Y_1^\dagger} {\bf Y_2} ] 
- {\bf Y_2} ( {4 \over 9} g_{B-L}^2 + 3 g_L^2 + 3 g_R^2
{16 \over 3} g_3^2) \}\nonumber\\
{d \over {dt}} {\bf Y^e_1}  & =  & { 1 \over {16 \pi^2}} \{ {\bf Y^e_1}
[ {\rm Tr} (3 {\bf Y_1^\dagger} {\bf Y_1}  + {\bf Y_1^{e\dagger}} {\bf Y^e_1} )
+ 4 {\bf Y_1^{e\dagger}} {\bf Y^e_1} + 
2 {\bf Y_2^{e\dagger}} {\bf Y^e_2} ]\nonumber\\
& + & {\bf Y^e_2} [ {\rm Tr} (3 {\bf Y_2^\dagger} {\bf Y_1}  
+ {\bf Y_2^{e\dagger}} {\bf Y^e_1} ) + 2 {\bf Y_2^{e\dagger}} {\bf Y^e_1} ] 
- {\bf Y^e_1} (  g_{B-L}^2 + 3 g_L^2 + 3 g_R^2) \}\nonumber\\
{d \over {dt}} {\bf Y^e_2}  & =  & { 1 \over {16 \pi^2}} \{ {\bf Y^e_2}
[ {\rm Tr} (3 {\bf Y_2^\dagger} {\bf Y_2}  + {\bf Y_2^{e\dagger}} {\bf Y^e_2} )
+ 2 {\bf Y_1^{e\dagger}} {\bf Y^e_1} + 
4 {\bf Y_2^{e\dagger}} {\bf Y^e_2} ]\nonumber\\
& + & {\bf Y^e_1} [ {\rm Tr} (3 {\bf Y_1^\dagger} {\bf Y_2}  
+ {\bf Y_1^{e\dagger}} {\bf Y^e_2} ) + 2 {\bf Y_1^{e\dagger}} {\bf Y^e_2} ] 
- {\bf Y^e_2} (  g_{B-L}^2 + 3 g_L^2 + 3 g_R^2) \}
\label{eq:twobid}
\ea

It is easy to see that the hermiticity of Yukawa couplings is
preserved throughout the running in the $SU(2)\times SU(2)\times U(1)_{B-L}$
phase ({\it i.e.} above $m_R$), as expected\footnote{For example note that
in the equation for ${\bf Y}_1$, we have a sum of terms 
${\bf Y_1} {\bf Y_2^\dagger} {\bf Y_2} +
{\bf Y_2} {\bf Y_2^\dagger} {\bf Y_1}$ which is hermitean if
$Y_1$ and $Y_2$ are hermitean.}.
This is in contrast 
to case a) where below $M_R$ running of 
matrices necessarily spoils hermiticity (both in the MSSM and the 4 Higgs 
doublet model), because then the LR symmetry is broken.

\vskip 0.3cm

\noindent{\bf APPENDIX D: Doublet-Doublet Splitting}

\vskip 0.6cm

In this Appendix, we show how a left-right symmetric theory 
with two bidoublets above the scale $M_R$ reduces to the MSSM with only
one pair of $(H_u,H_d)$. We will call this phenomenon doublet-doublet
splitting. The simplest way to achieve this is by a fine
tuning of the parameters of the superpotential  (\ref{eq:superpot})
involving the
$\phi_1$ and $\phi_2$ fields- i.e. $\mu_{ij}$. To make this explicit,
consider the part of the superpotential:
\begin{eqnarray}
W_{\phi}= \sum_{ij}{{1}\over{2}}\mu_{ij} Tr\phi^T_a \tau_2 \phi_b \tau_2
\end{eqnarray}
where the symbols $a,b$ go over 1,2. This leads to the following superpotential
in terms of the standard model doublets:
\begin{eqnarray}
W_{\phi}= \mu_{11} H_{u1}H_{d1} + \mu_{22} H_{u2} H_{d2} + \mu_{12} 
(H_{u1}H_{d2}+H_{u2}H_{d1})
\end{eqnarray}
Now it is clear that, if the parameters $\mu_{ab}$ are so chosen that we have
$\mu_{11}\mu_{22}-\mu^2_{12}=0$ and that each $\mu_{ij}$ are of order $v_R$,
then below the scale $v_R$, the model has only two standard model doublets
as in MSSM. The surviving doublets are then linear combinations of the
of the original four doublets in the theory. If however one wanted  ``pure"
doublets surviving below the $v_R$ scale (such as say, $H_{u1}$ and $H_{d2}$)
then one can use the superpotential of following type:
\begin{eqnarray}
W'_{\phi}= {{\lambda_{12}}\over{M_Pl}} {\rm Tr} \,\phi^T_1\tau_2\tau_i\phi_2 
Tr \Delta^c\tau_i\bar{\Delta}^c + \mu_{12} 
{\rm Tr} \phi^T_1 \tau_2\phi_2 \tau_2
\end{eqnarray}
In this case fine tuning of the parameters $\lambda_{12}v^2_R/M_{Pl}+\mu_{12}
=0$ leaves the pure low energy doublets $H_{u2}$ and $H_{d1}$.

%

\newpage
\textwidth 5.75in
\unitlength=1.00mm

\begin{figure}
\begin{picture}(33.66,182.34)
\put(30.00,120.00){\line(1,0){100.00}}
\multiput(60.00,120.00)(5,5){4}{\line(1,1){4.00}}
\multiput(100.00,120.00)(-5,5){4}{\line(-1,1){4.00}}
\put(40.00,123.00){\makebox(0,0)[rc]{$Q$}}
\put(120.00,123.00){\makebox(0,0)[rc]{$Q^c$}}
\put(74.00,124.00){\makebox(0,0)[rc]{$Q^c$}}
\put(90.00,124.00){\makebox(0,0)[rc]{$Q$}}
\put(83.00,142.00){\makebox(0,0)[rc]{$\mu_{ij}$}}
\put(83.00,116.00){\makebox(0,0)[rc]{$M_q^{(0)}$}}
\put(68.00,132.00){\makebox(0,0)[rc]{$\Phi_i$}}
\put(97.00,132.00){\makebox(0,0)[rc]{$\Phi_j$}}
\put(35.00,120.00){\vector(1,0){10}}
\put(115.00,120.00){\vector(-1,0){10}}
\put(80.00,120.00){\vector(-1,0){10}}
\put(80.00,120.00){\vector(1,0){10}}
\put(81.80,120.00){\makebox(0,0)[rc]{$\times$}}
\put(81.80,139.00){\makebox(0,0)[rc]{$\times$}}
\put(70,130){\vector(-1,-1){5}}
\put(90,130){\vector(1,-1){5}}
\end{picture}
\vspace{-8cm}
\caption{ Higgs contribution to one-loop 
calculation of $\bar{\Theta}$.}
\label{fig:higgs}
\end{figure}
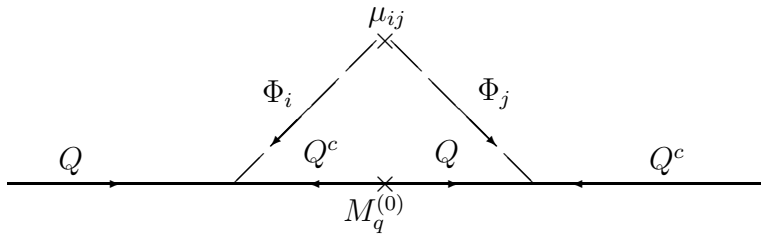

%

\newpage
\textwidth 5.75in
\unitlength=1.00mm
\thicklines
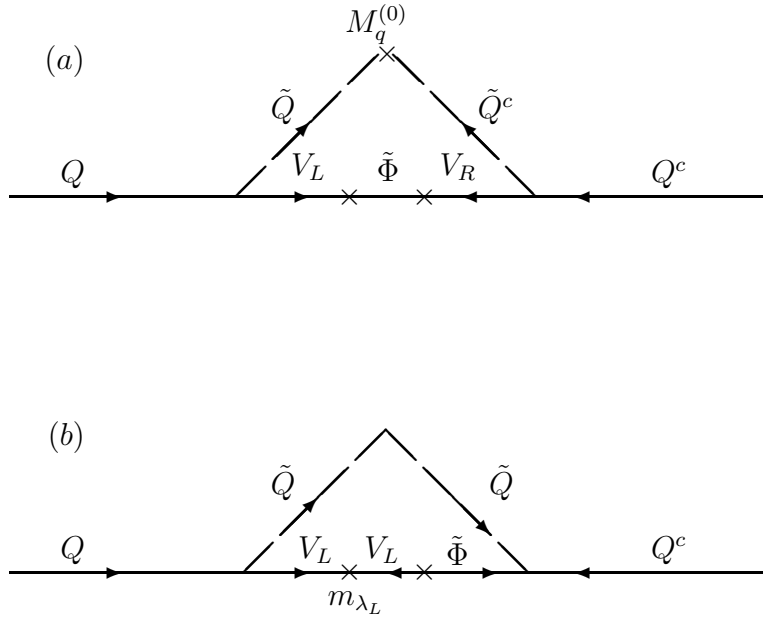
\begin{figure}
\begin{picture}(33.66,182.34)
\put(40.00,188.00){\makebox(0,0)[rc]{$(a)$}}
\put(30.00,170.00){\line(1,0){100.00}}
\multiput(60.00,170.00)(5,5){4}{\line(1,1){4.00}}
\multiput(100.00,170.00)(-5,5){4}{\line(-1,1){4.00}}
\put(83.00,193.00){\makebox(0,0)[rc]{$M_q^{(0)}$}}
\put(40.00,173.00){\makebox(0,0)[rc]{$Q$}}
\put(120.00,173.00){\makebox(0,0)[rc]{$Q^c$}}
\put(72.00,174.00){\makebox(0,0)[rc]{$V_L$}}
\put(81.80,174.00){\makebox(0,0)[rc]{$\tilde{\Phi}$}}
\put(92.00,174.00){\makebox(0,0)[rc]{$V_R$}}
\put(68.00,182.00){\makebox(0,0)[rc]{$\tilde{Q}$}}
\put(97.00,182.00){\makebox(0,0)[rc]{$\tilde{Q}^c$}}
\put(35.00,170.00){\vector(1,0){10}}
\put(115.00,170.00){\vector(-1,0){10}}
\put(60.00,170.00){\vector(1,0){10}}
\put(100.00,170.00){\vector(-1,0){10}}
\put(76.80,170.00){\makebox(0,0)[rc]{$\times$}}
\put(86.80,170.00){\makebox(0,0)[rc]{$\times$}}
\put(81.80,189.00){\makebox(0,0)[rc]{$\times$}}
\put(65,175){\vector(1,1){5}}
\put(95,175){\vector(-1,1){5}}
%
\put(40.00,138.00){\makebox(0,0)[rc]{$(b)$}}
\put(30.00,120.00){\line(1,0){100.00}}
\multiput(61.00,120.00)(5,5){4}{\line(1,1){4.00}}
\multiput(99.00,120.00)(-5,5){4}{\line(-1,1){4.00}}
\put(40.00,123.00){\makebox(0,0)[rc]{$Q$}}
\put(120.00,123.00){\makebox(0,0)[rc]{$Q^c$}}
\put(73.00,123.00){\makebox(0,0)[rc]{$V_L$}}
\put(91.00,123.00){\makebox(0,0)[rc]{$\tilde{\Phi}$}}
\put(81.80,123.00){\makebox(0,0)[rc]{$V_L$}}
\put(68.00,132.00){\makebox(0,0)[rc]{$\tilde{Q}$}}
\put(97.00,132.00){\makebox(0,0)[rc]{$\tilde{Q}$}}
\put(79.80,116.00){\makebox(0,0)[rc]{$m_{\lambda_L}$}}
\put(35.00,120.00){\vector(1,0){10}}
\put(115.00,120.00){\vector(-1,0){10}}
\put(60.00,120.00){\vector(1,0){10}}
\put(90.00,120.00){\vector(-1,0){10}}
\put(85.00,120.00){\vector(1,0){10}}
\put(76.80,120.00){\makebox(0,0)[rc]{$\times$}}
\put(86.80,120.00){\makebox(0,0)[rc]{$\times$}}
\put(66,125){\vector(1,1){5}}
\put(89,130){\vector(1,-1){5}}
\end{picture}
\vspace{-8cm}
\caption{Examples of gaugino contributions to one-loop 
calculation of $\bar{\Theta}$. $V_{L,R}$ are left and right gauginos, 
respectively. The gaugino mass $m_{\lambda_L}$ is in general complex.
There is an analogous graph to b) that involves right-handed gauginos.}
\label{fig:gaugino}
\end{figure}

\end{document}